\documentclass[conference,a4paper]{IEEEtran}
\def\ANONYMOUS{0}

\usepackage{graphicx} 

\usepackage{tikz}
\usetikzlibrary{graphs}
\usepackage{amsmath}
\usepackage{amssymb}
\usepackage[utf8]{inputenc}
\usepackage{babel}
\usepackage{amsthm}
\newtheorem{definition}{Definition}

\usepackage{hyperref}
\usepackage{float}
\usepackage{cuted} 
\usepackage[most]{tcolorbox}
\usepackage{amsmath}
\usepackage{xcolor}
\usepackage{enumitem}
\usepackage{algorithm}
\usepackage{algpseudocode}
\usepackage{msc}
\usepackage{booktabs,tabularx}

\newlist{steps}{enumerate}{1}
\setlist[steps]{label=Step \arabic*., leftmargin=*}

\usepackage{markdown}
\usepackage{standalone}

\usetikzlibrary{arrows.meta,positioning,shapes.misc,shapes.geometric,backgrounds}
\usetikzlibrary{calc}
\usetikzlibrary{shapes.symbols}

\newcommand{\tol}[1]{{\color{red} #1}} 

\newcommand{\GG}{\mathbb{G}}
\newcommand{\ZZ}{\mathbb{Z}}

\newcommand{\EEE}{\mathcal{E}}
\newcommand{\GGG}{\mathcal{G}}
\newcommand{\VVV}{\mathcal{V}}

\newcommand{\Commit}{\text{Commit}}
\newcommand{\MPEncode}{\text{MPEncode}}

\title{Topology-Hiding Connectivity-Assurance  \\for QKD Inter-Networking}

\ifnum\ANONYMOUS=0
    \author{\IEEEauthorblockN{Margherita Cozzolino}
    \IEEEauthorblockA{\textit{AIT Austrian Institute of Technology}\\
    Vienna, Austria \\
    {\tt margherita.cozzolino@ait.ac.at}}
    \and
    \IEEEauthorblockN{Stephan Krenn}
    \IEEEauthorblockA{\textit{AIT Austrian Institute of Technology}\\
    Vienna, Austria \\
    {\tt stephan.krenn@ait.ac.at}}
    \and
    \IEEEauthorblockN{Thomas Lorünser}
    \IEEEauthorblockA{\textit{AIT Austrian Institute of Technology}\\
    Vienna, Austria \\
    {\tt thomas.loruenser@ait.ac.at}}
    }
\else
    \author{\IEEEauthorblockN{Anonymous submission}
    \IEEEauthorblockA{QCNC 2026}}
\fi

\begin{document}

\maketitle

\begin{abstract}
 While QKD ensures information-theoretic security at the link level, real-world deployments depend on trusted repeaters, creating potential vulnerabilities. 
 In this paper, we thus introduce a topology-hiding connectivity assurance protocol to enhance trust in quantum key distribution (QKD) network infrastructures.
 Our protocol allows network providers to jointly prove the existence of a secure connection between endpoints without revealing internal topology details. 
 By extending graph-signature techniques to support multi-graphs and hidden endpoints, we enable zero-knowledge proofs of connectivity that ensure both soundness and topology hiding. 
 We further discuss how our approach can certify, e.g., multiple disjoint paths, supporting multi-path QKD scenarios. 
 This work bridges cryptographic assurance methods with the operational requirements of QKD networks, promoting verifiable and privacy-preserving inter-network connectivity.
 \end{abstract}
\begin{IEEEkeywords}
QKD networks, infrastructure certification, graph-signatures.
\end{IEEEkeywords}

\section{Introduction}

Quantum key distribution (QKD) emerges as an important technology to strengthen communication security in the quantum age.
The European Union envisions QKD as a building block to establish a quantum-safe communication infrastructure in Europe and is therefore encouraging the development of networks all over the continent \cite{Wissel:24}

Although QKD is information-theoretically secure, it has substantial limitations in real-world applications which have to be overcome to foster the widespread adoption of the technology.
On a network level, QKD can be considered as a symmetric key exchange primitive which is limited in distance and rate.
The limitations are intrinsic to the hardware and the quantum mechanic principle the security is based on.
To facilitate versatility in QKD, networks based on QKD links forward messages on a hop-by-hop basis and therefore introduce the concept of trusted intermediaries, called \emph{(trusted) repeaters} or relays.
These repeaters serve as fully trusted entities, and no security guarantees can be given in case of a single corrupted repeater on a transmission path.
However, this concept will remain at the heart of QKD networks \cite{Huttner_2022} for the upcoming future and is used in real-world testbeds \cite{elder_pan-european_qci_2024}.

\subsection{Motivation}\label{ss:motivation}
  Shifting away from the typical end-to-end security paradigm asks for new measures on the infrastructure level in order to obtain sufficiently high guarantees about the trustworthiness of the network operator and repeater nodes.

  However, while for instance in high-security contexts one may safely assume that stakeholders (e.g., governmental or military institutions) have information of the available national QKD network, this may neither be assumed for ``QKD-as-a-service'' scenarios offered, e.g., to industry, nor for trans-national communication, where data is being routed through potentially several countries.
  However, such trans-national and inter-network communications become increasingly important with the ongoing growth of the quantum communication infrastructure, e.g., in the European Union.

  Indeed, in the latter case it is not even possible for a user to reliably determine whether a connection between two nodes \emph{exists}, let alone perform a risk assessment based on the number of trusted repeaters or similar.
  
\begin{figure}[t]
  \centering
  \includegraphics[trim={0 0 12cm 0},clip,width=0.47\textwidth]{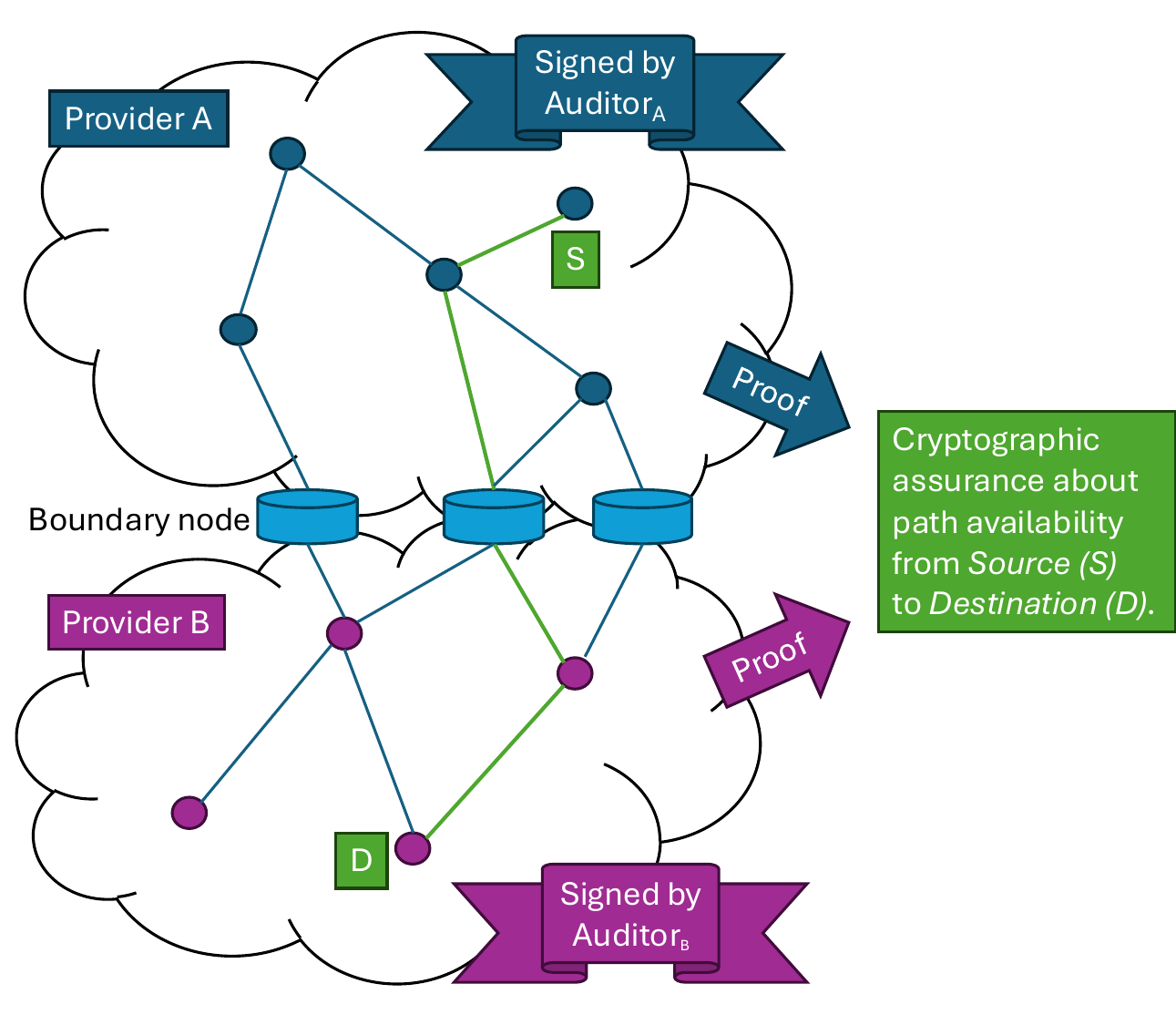}
  \caption{Users are able to obtain cryptographic assurance about inter-network path availability without revealing the underlying network topology.}
  \label{fig:qkd_inter_as}
\end{figure}

  Consider a scenario as depicted in Fig.~\ref{fig:qkd_inter_as}.
  There, a user wants to establish a secure link from the source ($S$) to the destination ($D$), each of which is under control of a different network provider.
  The two networks may have one or several \emph{boundary nodes}, which serve as connecting links between the networks.

  What would now be desirable is a formal and cryptographic proof jointly generated by the network providers, satisfying the following properties:
  \begin{itemize}
      \item \emph{Soundness.} The proof provides formal guarantees to the user that the two service providers can jointly offer a communication link between $S$ and $D$ with at most $\ell$ hops.
      \item \emph{Topology hiding.} The proof provably protects the business secrets of the service providers. 
      In particular, it neither reveals the actual path nor the utilized boundary node nor the number of hops performed within either network.
  \end{itemize}
  It is important to note that topology hiding (to the extent possible, i.e., the number of hops and the precise path within one's network) should also hold against the other involved network provider to enable horizontal collaboration of otherwise potentially competing providers.

\subsection{Related work}

Research is ongoing in reducing the trust required to put into trusted repeaters along a path and QKD network infrastructure in general.

Different approaches have been proposed in the literature.
One line of research is leveraging multi-path transmission protocols \cite{valbusa_relaxing_2025} to distribute the trust on multiple repeater nodes, even with different assumptions \cite{chen_quantum_2023}.
If data is securely split and sent among different disjoint paths, the adversary is required to compromise at least one repeater in every path, thus the work factor of the adversary is significantly increased.

Alternatively, hybrid cryptographic approaches combining quantum with classical cryptography are provided for relaxing the reliance on individual trusted nodes.
In \cite{brazaolavicario2025privacyenhancedqkdnetworks}, different approaches for reducing trust on QKD relay nodes are summarized and a new technique on fully-homomorphic cryptography was proposed.

However, besides the different approaches to the best of our knowledge, only little research was dedicated on the assurance of routing in QKD networks.

Only in \cite{franzoi_secure-key-forwarding_2025}, the authors provide a framework to assure the correctness of a chosen path for a message relayed in trusted repeater QKD networks.
The concept of QKD network assurance was developed further in \cite{todo-our-paper}, where attribute-based credentials have been introduced to assure properties of the used path to the user.
In their work they also introduce the idea of \emph{topology hiding}, which is important in the QKD context;
this property enables users to obtain cryptographically verifiable attestations about a network or transmission path, without revealing any additional information on the network topology itself.

In our work we build on the concept of topology certification as introduced in \cite{10.1007/978-3-662-47854-7_18}, where the idea of signatures on committed graphs was introduced, in a way that efficient zero-knowledge proofs about the graphs became possible.
Originally, this idea stems from research in attribute-based credentials and applies for certification of cloud infrastructure properties.
Such signatures allow one to prove statements about the signed graph, e.g., that two vertices are (or are not) connected by a sequence of edges.
A more efficient variant based on pairings was published later based on q-SDH signatures \cite{tan_q-sdh-based_2020}, which builds the basis for our new approach.
The RSA version have also be implemented \cite{sfyrakis_gsl_2020} and shows practical performance, but even better runtimes are expected for the q-SDH version, however, no q-SDH implementation has been proposed.

Although the concept of topology certification seems a promising candidate to increase assurance in QKD networks, it cannot be directly applied and requires substantial adaption and extension, which we are partially addressing in our work.

\subsection{Contribution and Technical Overview}
  In this paper we introduce a new concept which can help to establish trust into QKD network infrastructure by addressing the scenario mentioned before.

  The basic idea of our topology-hiding connectivity-assurance protocol is that a third party, the \emph{auditor}, checks and signs the network topology of a network provider.
  This may, e.g., be the national ministry of defense or any other governmental agency, and may be different for every network operator and country. 
  Akin to Gross et al.~\cite{10.1007/978-3-662-47854-7_18, tan_q-sdh-based_2020}, these signatures on network graphs can then be used to prove - in a zero-knowledge manner without disclosing any information than what is disclosed by the connectivity-claim itself - that different nodes in the network are indeed connected.
  Specifically, after agreeing on a boundary node to be used, the first service provider computes a proof for the path from $S$ to the boundary node, while the second computes a proof from the boundary node to $D$.
  
  However, while achieving soundness, using plain graph signatures as in~\cite{tan_q-sdh-based_2020,10.1007/978-3-662-47854-7_18} would not achieve privacy, as the number of hops per service provider as well as the boundary node are leaked.
  For this, we first extend \cite{tan_q-sdh-based_2020} to support multi-graphs:
  by adding loops (edges from a node to itself), we can then pad the path length to a predefined threshold, thereby hiding the actual number of hops within each domain.
  Secondly, we modify the proof system to support committed initial or terminal nodes, instead of revealing those in the plain:
  by this, the first provider can prove a (padded) path from $S$ to the \emph{committed} boundary node, and the second provider proves a (padded) path from the very same \emph{committed} boundary node to $D$. 
  
  Putting these enhancements of graph signatures together, we are able to propose an efficient and practical protocol QKD inter-networking, which achieves soundness and privacy as introduced in Sec.~\ref{ss:motivation}. Besides a basic protocol for scenarios as described above, we also discuss options how the protocol may be extended in order to not only prove availability of a single path from $S$ to $D$, but of $k$ \emph{disjoint} paths in the spirit of multi-path QKD.


\section{Preliminaries}
    Let $\GG_1,\GG_2,\GG_T$ be three cyclic groups of prime order $p$, based on an elliptic curve whose bilinear pairing is $\text{e}:\GG_1 \times \GG_2 \to \GG_T$. 
    
    Further, let MPEncode be as defined in \cite{cryptoeprint:2020/1403}, where MPEncode$:\ZZ^n_p \to \ZZ^{n+1}_p$ converts a set of messages into coefficients of a monic polynomial of degree $n+1$.

\subsection{Cryptographic Definitions}
    The following definitions are introduced to support the security proofs of the schemes presented later in this paper.
    \begin{definition}[\cite{cryptoeprint:2020/1403}]
        An algorithm $\mathcal{C}$ is said to ($t_{codlog},\varepsilon_{codlog}$)-break the co-DLOG assumption if $\mathcal{C}$ runs in time at most $t_{codlog}$ and 
        $$\text{PR}[x \in \ZZ_p : \mathcal{C}(g_1,g_1^x \in \GG_1, g_2, g_2^x \in \GG_2)=x]\ge \varepsilon_{dlog}.$$ 
    \end{definition}
    \begin{definition}[\cite{cryptoeprint:2020/1403}]
        An algorithm $\mathcal{C}$ is said to ($t_{sdh},\varepsilon_{sdh}$)-break the SDH assumption if $\mathcal{C}$ runs in time at most $t_{sdh}$ and 
        \begin{align*}
          \text{PR}[&x \in \ZZ_p , c\in \ZZ_p\setminus \{-x\}: \\
             &\mathcal{C}(g_1,g_1^x \dots,g_1^{x^q} g_2, g_2^x )=(g_1^{\frac{1}{x+c}},c))]\ge \varepsilon_{cdh}.
        \end{align*}
    \end{definition}

\subsection{Pedersen Commitment}
\label{subsec:Pedersen}

    A commitment scheme is a protocol that allows one party (the committer) to fix a chosen value without disclosing it, with the ability to reveal it later in a verifiable way.
    \begin{definition}
        A commitment scheme is perfectly hiding if, for any pair of messages $m_0$ and $m_1$, the distributions of the commitments are identical. 
    \end{definition}
    \begin{definition}
        A commitment scheme is computationally binding if no efficient (i.e., PPT) committer can open a commitment to two different messages with more than negligible probability.
    \end{definition}

Let $\GG$ be a group of prime order $p$, and let $g,h \gets \GG$. A Pedersen commitment~\cite{DBLP:conf/crypto/Pedersen91} of a message $m\in\ZZ_p$ is computed as 
$$C=\Commit(pk,m,r)=g^mh^r \quad\text{for}\quad r\gets \ZZ^*_p,$$
where $r$ is the opening value. The Pedersen commitment scheme is perfectly hiding and computationally binding under the discrete logarithm assumption in $\GG$. 

We will employ it within the statement connected to hide the initial and terminal nodes of the path.

\subsection{Zero-Knowledge Notation}
    A zero knowledge proof ($ZKP$) is a cryptographic protocol that allows a prover to convince a verifier that they possess certain information, without revealing anything about that information itself. This technique is particularly useful for preserving data privacy and ensuring that sensitive information remains hidden. A zero knowledge proof can be realized as interactive protocol between the prover and the verifier, or as a non-interactive scheme under specific assumptions. 
    
    We use the following notation introduced by Camenisch and Stadler~\cite{DBLP:conf/crypto/CamenischS97}:
    $$PK\{ (\alpha, \beta, \delta): y=g^\alpha h^\beta \ \wedge \ \Tilde{y}=\Tilde{g}^\alpha\Tilde{h}^\delta \ \wedge \ (u \le \alpha \le v) \}.$$
    This denotes a zero knowledge proof of knowledge of integers $\alpha, \beta, \delta$ such that $y=g^\alpha h^\beta$, $\Tilde{y}=\Tilde{g}^\alpha\Tilde{h}^\delta$ and $u \le \alpha \le v$ hold, where $y,g,h,\Tilde{y},\Tilde{g}$ and $\Tilde{h}$ are elements of some groups $G=\langle g\rangle=\langle h\rangle$ and $\Tilde{G}=\langle \Tilde{g}\rangle=\langle \Tilde{h}\rangle$. According to the standard convention, quantities denoted by Greek letters correspond to values whose knowledge is being proven, while all remaining variables are assumed to be known by the verifier.

\subsection{Multigraph MoniPoly Encoding}\label{subsec:param}
Tan et al.~\cite{cryptoeprint:2020/1403} proposed a graph signature scheme based on the MoniPoly attribute-based credential system~\cite{monipoly-asiacrypt}.
However, while they show how to encode graph data structures, their work focuses on simple graphs, lacking the functionality of loops as needed in our construction for padding the path length. In the following, we thus now show how this encoding can be extended to multigraphs. We consider multigraphs $\GGG=(\VVV,\EEE)$ over finite sets of vertices and multisets edges, with undirected edges and finite sets of vertex and edge labels. In this work, we focus on multigraphs where certain nodes possess multiple loops, while configurations involving multiple edges between distinct nodes are not explicitly considered as not required for our purpose. 
\begin{itemize}
    \item $\VVV$ is the finite set of vertices,
    \item $\EEE \subseteq \VVV \times \VVV$ is the finite multiset of edges,
    \item $\Xi_\VVV, \Xi_\mathcal{L}$ are the vertex identifier universe and labels universe,
    \item $f_{\VVV}: \VVV \to \mathcal{P}(\mathcal{L}_{\VVV})$ are the labels of a given vertex and $f_{\EEE}: \EEE \to \mathcal{P}(\mathcal{L}_{\EEE})$ are the labels of a given edge,
    \item $n=|\VVV|, \ m=|\EEE|$ are the numbers of vertices and edges, $L=|G|=|\VVV|+|\EEE|$ is the dimension of the graph.
\end{itemize}

For the encoding, we now follow the lines of Tan et al.~\cite{cryptoeprint:2020/1403}, by generating (based on a secret value $x'$) the appropriate parameters ${a}_{0_0}=g_1\in \GG_1,$ ${X}_{0_0}=g_2\in \GG_2$, $\{ a_{0_k}=a_{0_0}^{x'^k}, X_{0_k}=X_{0_0}^{x'^k} \}_{k=1}^n$, and $\{ \{ {a_{i_k}}, {X_{i_k}} \}^L_{i=1}\}^n_{k=0}$; 
for details on the generation, we refer to the original literature.

For every vertex $i$, we define $\VVV_i= \{ i, f_\VVV(i)\}$ ($|\VVV_i|=n_i$) and for every edge $(i,j)$, not loop, $\EEE_{(i,j)}= \{ i,j,f_\EEE(i,j)\}$ ($|\EEE_{(i,j)}|=n_{(i,j)}$). 
Now, in contrast to the original work, since nodes can have many loops, we need to avoid ambiguity in case that two loops have the same labels.
We thus extend the label of each loop by a counter to ensure unique identifiers, resulting in $\EEE_{(i,i)}=\{ i,i,f_\EEE(i,i),\text{counter}\}$, providing a well-formed graph encoding in the sense of Tan et al.~\cite{cryptoeprint:2020/1403}.

    Let now $\{ {\text{m}_{i_k}}\}=\MPEncode(\VVV_i)$ and $\{ {\text{m}_{{(i,j)}_k}}\}=\MPEncode(\EEE_{(i,j)})$. 
    The encoding of a vertex $i$ and an edge $(i,j) $ are then given by, respectively,
    \begin{alignat*}{2}
\text{Encoding vertex }i  & = \prod_{k=1}^{n_i} {{a}_i}_k^{\text{m}_{i_k}}, \\
\text{Encoding edge }(i,j) &  = \prod_{k=1}^{n_{(i,j)}} {{a}_{(i,j)}}_k^{{\text{m}_{(i,j)}}_k}.
\end{alignat*}

\subsection{Multigraph MoniPoly Commitment}
\label{subsec: commitment}
    In this section, we recap the process by which the multigraph $\GGG=(\VVV,\EEE)$ is committed using a variant of the MoniPoly set commitment \cite{cryptoeprint:2020/1403}.
    Using the same parameters as before, each vertex $i\in \VVV $ is committed as
        $$C_i= 
        (\prod_{k=1}^{n_i} {{a}_i}_k^{{\text{m}_{i_k}}})^{o_i}$$
        while each edge $(i,j)\in \EEE$ is committed as 
        \begin{align*}
            C_{(i,j)}
            &=(\prod_{k=1}^{n_{(i,j)}} {{a}_{(i,j)}}_k^{{\text{m}_{(i,j)}}_k})^{o_{(i,j)}}
        \end{align*}
        where $o_i$ and ${o_{(i,j)}}$ are random opening values in ${\ZZ_p^*}$ and $\{ {\text{m}_{i_k}}\}=\MPEncode(\VVV_i) \ \forall i \in \VVV$ and $\{ {\text{m}_{{(i,j)}_k}}\}=\MPEncode(\EEE_{(i,j)}) \ \forall {(i,j)} \in \EEE$. Note that all loops commitments are distinct. If a node $i$ has more than one loop, each of them is assigned a unique commitment, since the corresponding encodings are all different. 
        The commitment $C$ of the graph $\GGG$ is the product of the vertices commitment $C_i$ and edges commitment $C_{(i,j)}$:
        $$C=\prod_{i \in \VVV} C_i \cdot \prod_{(i,j) \in \EEE} C_{(i,j)} $$

        The randomization of $C$ is perfectly hiding, and the Multigraph MoniPoly Commitment is binding if the co-DLOG problem is hard \cite{cryptoeprint:2020/1403}. 

\subsection{Camenisch-Lysyanskaya SDH Signature Scheme} 
In this section, we briefly recall the CL-SDH signature scheme~\cite{DBLP:conf/crypto/CamenischL04} as used in~\cite{cryptoeprint:2020/1403,cryptoeprint:2023/1181}. 
This signature scheme serves as a fundamental building block for the multigraph signing scheme presented in the following section. 
The scheme is defined over the message space $\mathcal{M}$ and consists of the following algorithms: 

\begin{itemize}
    \item \emph{KeyGen$(1^k)$} defines the necessary parameters as before (based on some secret $x'$, for details cf. \cite{cryptoeprint:2020/1403}) and chooses $x\in \ZZ^*_p$, $a,b,c\in \GG_1,g_2\in \GG_2$. 
    Furthermore, it sets $X=g_2^x$ and the outputs the parameters along with $X$ as public key, as well as the secret key $sk=(x,x').$
    \item \emph{Sign$(pk,sk,\{m_1,\dots,m_2 \})$} chooses the random values $s,t\in \ZZ^*_p$ to compute $v=(a_0^{\prod_{i=0}^n (x'+m_i)}b^sc)^{\frac{1}{x+t}}$ and output the signature $\sigma=(t,s,v)$. 
    \item \emph{Verify$(pk,\sigma,\{ m_1\dots m_n\})$} computes $\{ \text{m}_i\}_{i=0}^n=\MPEncode(\{ m_i\}_{i=1}^n)$ and output 1 if e$(v,X)=$e$(\prod_{i=0}^n a_i^{\text{m}_i}b^s c v^{-t},g_2 )$ holds, otherwise output 0. 
\end{itemize} 

    The CL-SDH signature is strongly existential unforgeable against chosen message attack in the standard model if the SDH problem is intractable~\cite{DBLP:conf/crypto/CamenischL04}.

\subsection{Signature Scheme on Committed Multigraph}
\label{subsec:sig}
Obtaining a signature on a committed multigraph is now an interactive protocol between the signer and the holder of the graph as detailed in \cite{cryptoeprint:2020/1403}.
  We omit giving full details here, and just note that the extension from signatures on graphs to multigraphs is a straightforward adaptation.

  At the end of the interaction the graph holder now obtains a signature of the form $\sigma=(t,s,v,\GGG)$
where $\GGG=(\VVV,\EEE), s,t\in \ZZ^*_p$ and
\begin{align*}
v&= (h^{sk_U}
\underbrace{ \dots a_{i_0}^{(x'+i)\prod_{k \in f_\VVV(i)} (x'+k)} \dots}_{\forall \ \text{vertices} \ i\in \VVV}\\
&\underbrace{\dots a_{(i,j)_0}^{(x'+i)(x'+j)\prod_{k \in f_\VVV((i,j))} (x'+k)} \dots}_{\forall \ \text{edges} \ (i,j)\in \EEE} b^sc)^{\frac{1}{x+t}}.
\end{align*}
Here, $h,b,c$ are public parameters and $x,x'$ are secret values, generated by the signer, while $sk_U$ is the secret key of the graph holder, required to prevent impersonation attacks.

The construction of $v$ is based on the CL-SDH signature scheme, defined in the previous section, on the multigraph encoding. The symbol $v$ shown above does not represent the exact output of the signature, but rather provides a structural intuition of it. This observation illustrates that extending the signature to multigraphs is both feasible and conceptually straightforward. The only distinction between the signature on a graph and that on a multigraph lies in the computation of $v$: in the former, $v$ is derived from the encoding and commitment of the graph, whereas in the latter it is computed using the encoding and commitment of the multigraph, which introduces additional values, one for each loop added to the nodes, absent in the graph case. 

In any case, the parameters chosen by the signer, $h,b,c,x,x'$ do not affect the feasibility of applying the signature to multigraphs. 

The graph signature scheme can be shown to satisfy impersonation resilience under concurrent attacks (IMP-ACA) (that guarantees that a malicious prover cannot deceive the verifier into believing that they possess a valid signature, nor can they make the verifier accept false statements), graph unlinkability under active and concurrent attacks (GUNL-ACA) (that ensures that an adversary gains no information about the provider's signed graph $\GGG$) and protocol unlinkability under active and concurrent attacks (PUNL-ACA) (that ensures that an adversary should not be able to learn anything about the hidden graph $\GGG$), initially introduced in the MoniPoly ABC framework \cite{cryptoeprint:2020/587}. The security properties of the signature scheme are formally proven in \cite{cryptoeprint:2020/1403}. 

\section{A Blackbox Construction}
    We are now ready to present a high-level protocol for proving availability of an inter-network path between networks provided by providers $A$ and $B$. 
    
    Specifically, let $A$ be the owner of a network with structure $\GGG_A=(\VVV_A,\EEE_A)$, and  $B$ the owner of the graph $\GGG_B=(\VVV_B,\EEE_B)$. These graphs are secret, only the respective provider knows how they are composed. A user $\mathcal{C}$ wishes to obtain a path from a source $S\in \VVV_A$ to a destination $D\in \VVV_B$. 
    
    The protocol consists of two phases, where the first only needs to be executed once when setting up the system, while the second phase needs to be executed for a specific path whose existence is to be assured.
    
  \textbf{Phase 1}: \textit{Network Certification}. In this phase, each provider, say $A$ and $B$, obtains from its respective certification authority ($CA$) - e.g., the national Ministry of Defense or any other trusted authority - a signature on their network topology. 
  
  We again stress that each network provider may obtain a signature from a different authority - there are no technical constraints requiring the same signing key to be used for different networks.
    However, it is important that adjacent network providers agree on the labeling (i.e., naming conventions) only for the boundary nodes;
    no inter-network naming conventions for internal nodes are needed. \\

    \textbf{Phase 2}: \textit{Proof of Path Existence}. 
    This phase consists of the following steps:
    \begin{itemize}
        \item Step 1: \textit{Finding the paths}. Providers $A$ and $B$ determine the cheapest paths in their original graphs (this can be, for instance, defined as the minimum number of edges traversed to connect two nodes): the first one searches for the cheapest paths from $S$ to every boundary node, the second one does the same from the node $D$.
        
        \item Step 2: \textit{Boundary node agreement}. Providers $A$ and $B$ execute a protocol to minimize the total connection cost and agree on the boundary node $BN$ to be used as the connection point. This process could be done, e.g., by using a secure multi-party computation (aka MPC) protocol~\cite{DBLP:conf/focs/Yao86,DBLP:journals/ftsec/EvansKR18}.
        
        \item Step 3: \textit{Commitment agreement}. The provider $A$ commits to the selected boundary node $BN$ and sends the corresponding opening value to the provider of the multigraph $\GGG'_B$ using a secure communication channel.
        
        \item Step 4: \textit{Path existence in multigraph} $\GGG'_A$. The provider $A$ proves to the customer  $\mathcal{C}$ the existence of a path in the signed multigraph $\GGG'_A$ from node $S$ to the hidden bounary node $BN$, using the connected statement showed in Sec.~\ref{subsec: connected}. 
        On a high level, this corresponds to computing a zero-knowledge proof showing that:
\begin{itemize}
            \item $A$ holds a signature on $\GGG_A$ issued by the responsible auditor,
            \item $\GGG_A$ contains a path from $S$ to some boundary node $BN$ via at most $\ell_A$ edges, and
            \item $BN$ is contained in the commitment $C_{BN}$.
\end{itemize}
        
        \item Step 5: \textit{Path existence in multigraph} $\GGG'_B$. Provider $B$ computes a similar proof, showing that:
\begin{itemize}
            \item $B$ holds a signature on $\GGG_B$ issued by the responsible auditor,
            \item $\GGG_B$ contains a path from some boundary node $BN$ to $D$ via at most $\ell_B$ edges, and
            \item $BN$ is contained in the same commitment $C_{BN}$ as for provider $A$.
\end{itemize}
    \end{itemize}

Figure~\ref{fig:message-flow-onecol} illustrates the protocol’s message flow.

\begin{figure}[H]
\centering
\resizebox{\columnwidth}{!}{%
\begin{tikzpicture}[
    x=0.95cm, y=0.78cm, >=Latex,
    lifeline/.style={draw=black!60, dashed},
    phase/.style={draw, rounded corners, fill=gray!10, inner sep=2pt, font=\footnotesize},
    msg/.style={-Latex, line width=0.4pt},
    note/.style={font=\scriptsize},
    myfont/.style={font=\scriptsize}
]
\node[myfont] (caA) at (0,0)  {$CA_A$};
\node[myfont] (A)   at (3,0)  {Provider $A$};
\node[myfont] (B)   at (7,0)  {Provider $B$};
\node[myfont] (caB) at (10,0) {$CA_B$};
\node[myfont] (C)   at (13,0) {Customer $\mathcal{C}$};

\draw[lifeline] (caA) -- ++(0,-14);
\draw[lifeline] (A)   -- ++(0,-14);
\draw[lifeline] (B)   -- ++(0,-14);
\draw[lifeline] (caB) -- ++(0,-14);
\draw[lifeline] (C)   -- ++(0,-14);

\node[phase] at (6.5,-1.6) {Phase 1: Network Certification};

\draw[msg] (A)   -- node[above,sloped, note, yshift=2pt]{Sig req on $\mathcal{G}_A$} (caA |- 0,-2.3);
\draw[msg] (caA |- 0,-3.0) -- node[above,sloped, note, yshift=2pt]{$\sigma_A$ on $\mathcal{G}'_A$} (A |- 3,-3.0);

\draw[msg] (B)   -- node[above,sloped, note, yshift=2pt]{Sig req on $\mathcal{G}_B$} (caB |- 10,-2.3);
\draw[msg] (caB |- 10,-3.4) -- node[above,sloped, note, yshift=2pt]{$\sigma_B$ on $\mathcal{G}'_B$} (B |- 7,-3.4);

\node[phase] at (6.5,-4.3) {Phase 2: Proof of Path Existence};

\node[note, anchor=west] at (A |- 3,-5.3) {$A$: cheapest paths $S\to \mathcal{B}_A$};
\node[note, anchor=west] at (B |- 7,-5.9) {$B$: cheapest paths $D\to \mathcal{B}_B$};

\draw[msg] (A |- 3,-6.9) -- node[above,sloped, note, yshift=2pt]{MPC inputs (costs/cands)} (B |- 7,-6.9);
\draw[msg] (B |- 7,-7.6) -- node[above,sloped, note, yshift=2pt]{MPC inputs (costs/cands)} (A |- 3,-7.6);
\node[note] at (5,-8.3) {MPC selects $BN$};

\node[note, anchor=west] at (A |- 3,-9.3) {$C_{BN_1},C_{BN_2}$};
\draw[msg] (A |- 3,-10.0) -- node[above,sloped, note, yshift=2pt]{Openings of $C_{BN_1},C_{BN_2}$ (secure)} (B |- 7,-10.0);
\node[note, anchor=west] at (B |- 7,-10.7) {Verify opening of $C_{BN}$};

\draw[msg] (A |- 3,-11.9) -- node[above,sloped, note, yshift=2pt]{Connected($S,BN,\ell_A$) with hidden $BN$} (C |- 13,-11.9);
\draw[msg] (B |- 7,-12.9) -- node[above,sloped, note, yshift=2pt]{Connected($BN,D,\ell_B$) with hidden $BN$} (C |- 13,-12.9);

\end{tikzpicture}%
}
\caption{}
\label{fig:message-flow-onecol}
\end{figure}

\subsection{Phase 1} 

\begin{figure}
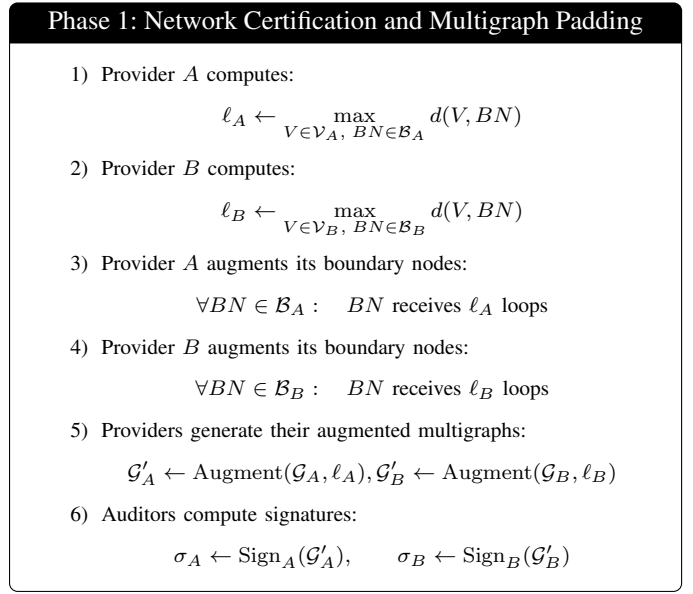

\begin{tcolorbox}[
title=Phase 1: Network Certification and Multigraph Padding,
fontupper=\footnotesize,
colback=white,
colframe=black,
boxrule=0.4pt
]


\begin{enumerate}
    \item Provider $A$ computes:
    \[
    \ell_A \gets 
    \max_{V \in \VVV_A,\; BN \in \mathcal{B}_A} d(V,BN)
    \]

    \item Provider $B$ computes:
    \[
    \ell_B \gets 
    \max_{V \in \VVV_B,\; BN \in \mathcal{B}_B} d(V,BN)
    \]

    \item Provider $A$ augments its boundary nodes:
    \[
    \forall BN \in \mathcal{B}_A:\quad BN \text{ receives } \ell_A \text{ loops}
    \]

    \item Provider $B$ augments its boundary nodes:
    \[
    \forall BN \in \mathcal{B}_B:\quad BN \text{ receives } \ell_B \text{ loops}
    \]

    \item Providers generate their augmented multigraphs:
    $$
    \GGG'_A \gets \mathrm{Augment}(\GGG_A,\ell_A), \GGG'_B \gets \mathrm{Augment}(\GGG_B,\ell_B)
    $$

    \item Auditors compute signatures:
    \[
    \sigma_A \gets \mathrm{Sign}_A(\GGG'_A),
    \qquad
    \sigma_B \gets \mathrm{Sign}_B(\GGG'_B)
    \]
\end{enumerate}
\end{tcolorbox}
\caption{Summary of Phase 1}\label{fig:phase_1}
\end{figure}

Phase 1, also detailed in Fig.~\ref{fig:phase_1}, prepares certified, topology-hiding inputs for the connectivity proofs.
Each provider $X\in\{A,B\}$ computes the padding target
$\ell_X=\max_{v\in\VVV_X,\; b\in\mathcal{B}_X} d(v,b)$, i.e., the longest
shortest-path distance from any internal node to any boundary node in its own
network. Provider $X$ then adds $\ell_X$ self-loops to every boundary node,
obtaining the padded multigraph $\GGG'_X$. This loop inflation allows any real path to be padded to length $\ell_X$, so the subsequent proof length does not leak internal distances. 
That is, in the presentation they will allow us to conceal the actual path length within graph $\GGG_A$, from node $S$ to the boundary node $BN$ in multigraph $\GGG_A '$, and similarly the path length from node $BN$ to the node $D$ in multigraph $\GGG_B'$, and thus achieve topology hiding.

Finally, the corresponding auditor issues a signature $\sigma_X$ on $\GGG'_X$, binding the
augmented topology to a certified identity. The outputs $(\GGG'_A,\GGG'_B,
\sigma_A,\sigma_B)$ and targets $(\ell_A,\ell_B)$ are then consumed in
Phase~2.

\subsection{Phase 2}\label{subsec:phase3}

\begin{figure}
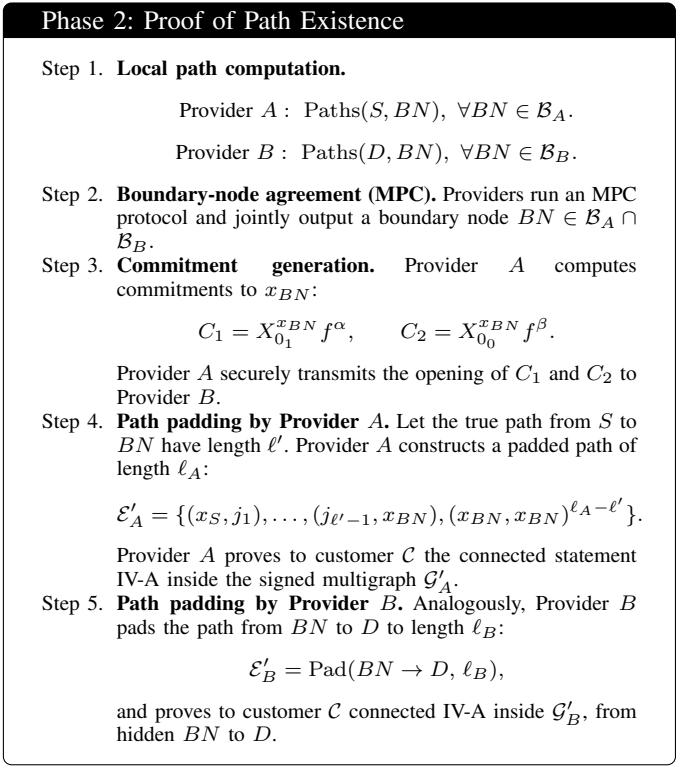

\begin{tcolorbox}[
title=Phase 2: Proof of Path Existence,
fontupper=\footnotesize,
colback=white,
colframe=black,
boxrule=0.4pt
]



\begin{steps}

    \item \textbf{Local path computation.}
    \[
    \text{Provider } A:\ \mathrm{Paths}(S, BN),\ \forall BN \in \mathcal{B}_A.
    \]
    \[
    \text{Provider } B:\ \mathrm{Paths}(D, BN),\ \forall BN \in \mathcal{B}_B.
    \]

    \item \textbf{Boundary-node agreement (MPC).}
    Providers run an MPC protocol and jointly output a boundary node
    \(
    BN \in \mathcal{B}_A \cap \mathcal{B}_B.
    \)

    \item \textbf{Commitment generation.}
    Provider $A$ computes commitments to $x_{BN}$:
    \[
    C_1 = X^{x_{BN}}_{0_1} f^\alpha,
    \qquad
    C_2 = X^{x_{BN}}_{0_0} f^\beta.
    \]
    Provider $A$ securely transmits the opening of $C_1$ and $C_2$ to Provider $B$.

    \item \textbf{Path padding by Provider $A$.}
    Let the true path from $S$ to $BN$ have length $\ell'$.  
    Provider $A$ constructs a padded path of length $\ell_A$:
    \[
    \EEE'_A = \{(x_S,j_1),\dots,(j_{\ell'-1},x_{BN}),
    (x_{BN},x_{BN})^{\ell_A-\ell'}\}.
    \]
    Provider $A$ proves to customer $\mathcal{C}$ the connected statement \ref{subsec: connected} inside the signed multigraph $\GGG'_A$.

    \item \textbf{Path padding by Provider $B$.}
    Analogously, Provider $B$ pads the path from $BN$ to $D$ to length $\ell_B$:
    \[
    \EEE'_B = \mathrm{Pad}(BN \to D,\, \ell_B),
    \]
    and proves to customer $\mathcal{C}$ connected \ref{subsec: connected} inside $\GGG'_B$, from hidden $BN$ to $D$.

\end{steps}


\end{tcolorbox}
\caption{Summary of Phase 2}\label{fig:phase_2}
\end{figure}

In Phase~2, cf. Fig.~\ref{fig:phase_2}, the providers locally compute candidate paths and use MPC to agree on a
boundary node $BN$ without revealing preferences or costs. Provider $A$ commits to
$x_{BN}$ (and shares the openings with $B$), so that both proofs bind to the same
hidden boundary node. Each provider then pads its path to the certified targets
$\ell_A,\ell_B$, and proves to the customer $\mathcal{C}$ a connectivity
statement in the signed multigraph ($\GGG'_A$ for $A$, $\GGG'_B$ for $B$):
$A$ proves availability of a path from $S$ to $BN$ with length $\ell_A$ and membership of $BN$ in the
commitment; symmetrically, while $B$ proves availability of a path from $BN$ to $D$ with length $\ell_B$
against the same commitment. The verifier accepts only if both ZK proofs verify,
establishing an end-to-end path from $S$ to $D$ while keeping internal topology hidden.

\section{Connected(\texorpdfstring{$i^*,j^*,\ell$}{i*,j*,l}) with hidden \texorpdfstring{$i^*,j^*,\ell$}{i*,j*,l}}

Connected($i^*,j^*,\ell$) represents the central concept of this work.
The Connected($i^*,j^*,\ell$) statement, as defined in the paper \cite{cryptoeprint:2020/1403}, enables the prover, who holds a committed and signed graph $\GGG=(\VVV,\EEE)$, to demonstrate to a verifier the existence of a path between nodes $i^*$ and $j^*$ of length $\ell$, where the length is defined as the number of edges connecting the two nodes.
Building on this idea, we propose an approach that allows proving  the existence of a path between two hidden nodes, $i^*$ and $j^*$, without explicitly revealing their identity and the exact path length. 

The proof can also be applied in the scenario where only one of the two nodes is hidden. However, in the following, we focus on the more general case in which both nodes remain hidden, thus proving a general solution applicable to other settings as well. In the setting described above, provider $A$ proves the existence of a path of length $\ell_A$ from the public node $S$ (denoted here as $i^*$) to the hidden boundary node $BN$ (denoted as $j^*$), while provider $B$ proves the existence of a path of length $\ell_B$ from the hidden node $BN$ (here $i^*$) to the public node $D$ (here $j^*$). 

The case of proving a path between a public and a hidden node slightly differs from the general setting; further details are provided at the end of the next subsection. 

We exploit the structure of a multigraph since this construction allows us to increase the original path length to reach a desired target value, while preserving the correctness and consistency of the proof.

\subsection{Multigraph Connection Proof}\label{subsec: connected}

    In this section, we present the zero-knowledge proof demonstrating the existence of a path between two hidden nodes, $i^*$ and $j^*$, without disclosing its exact length.
    All newly introduced values will be described in the following section, providing a clear understanding of the overall functioning of the proof and enabling the reader to fully grasp its underlying mechanisms. 
    
    Suppose that the path whose existence the prover wishes to demonstrate is 
    $$\EEE'=\{ (x^*,j_1),(j_1,j_2),\dots, (j_{\ell'-1},j^*),(j^*,j^*)^{\ell-\ell'} \}.$$
    Let $x_{i^*}$ and $x_{j^*}$ the identifiers of the vertices $i^*$ and $j^*$. 
    
    The prover selects random $\{ r_i, r_{(i,j)}, r_{(x_{i^*},x_{j^*})} \} \in \ZZ^*_p$ and computes Pedersen commitments as follows:     
    \begin{align*}
        C_1&=X^{x_{i^*}}_{0_1}\cdot f^{\alpha}, \\
        C_2&=X^{x_{i^*}}_{0_0}\cdot f^{\beta}\,,\\
        C_3&=X^{x_{j^*}}_{0_1}\cdot f^{\delta}, \text{ and} \\ 
        C_4&=X^{x_{j^*}}_{0_0}\cdot f^{\mu}\,,\\
    \end{align*}
    
    where $\alpha,\beta,\delta$ and $\mu$ are opening random values in $\ZZ^*_p$. The proof of knowledge is the following: 

 \begin{align*}
             & PK \bigg\{ (\forall i \in \VVV : \varepsilon_{i_0}, \varepsilon_{i_1}), (\forall (i,j) \in \EEE \setminus \EEE': {\varepsilon_{(i,j)}}_0,{\varepsilon_{(i,j)}}_1), \\
             & \{ {\varepsilon_{l,0}}\} ^{\ell  -1}_{l=2}, 
             \{ {\varepsilon_{l,1}}\} ^{\ell  -1}_{l=1}, \{ {\varepsilon_{l,2}}\} ^{\ell}_{l=1}, \\
             &\zeta, \rho, \omega, \tau, \gamma, \alpha , \beta, q, x_{i^*}, \delta, \mu, d, x_{j^*}, z): \\
             &\prod_{i \in \VVV} \text{e}(W'_i, X^{\varepsilon_{i_1}}_{0_1} X^{\varepsilon_{i_0}}_{0_0}) \prod_{(i,j)\in \EEE \setminus \EEE'} \text{e}(W'_{(i,j)}, X^{{\varepsilon_{(i,j)}}_1}_{0_1} X^{{\varepsilon_{(i,j)}}_0}_{0_0}) \cdot \\
             & \text{e}(W'_1, X^{\varepsilon_{1,2}}_{0_2} C_1^{\varepsilon_{1,2}} X^{\varepsilon_{1,1}}_{0_1} C_2^{\varepsilon_{1,1}}f^q) \text{e}(W'_2, X^{{\varepsilon_{2,2}}}_{0_2} X^{{\varepsilon_{1,1}}}_{0_1} X^{{\varepsilon_{2,1}}}_{0_1} X^{{\varepsilon_{2,0}}}_{0_0}) \dots \\
             & \dots \text{e}(W'_{\ell' -1}, X^{{\varepsilon_{\ell' -1,2}}}_{0_2} X^{{\varepsilon_{\ell' -2,1}}}_{0_1} X^{{\varepsilon_{\ell' -1,1}}}_{0_1} X^{{\varepsilon_{\ell' -1,0}}}_{0_0}) \cdot \\   
            & \text{e}(W'_{\ell'}, X^{{\varepsilon_{\ell' ,2}}}_{0_2} X^{{\varepsilon_{\ell' -1,1}}}_{0_1} X^{{\varepsilon_{\ell' ,1}}}_{0_1}  X^{{\varepsilon_{\ell',0}}}_{0_0}) \\
            & \text{e}(W'_{\ell'+1}, X^{{\varepsilon_{\ell'+1 ,2}}}_{0_2} X^{{\varepsilon_{\ell',1}}}_{0_1} X^{{\varepsilon_{\ell'+1 ,1}}}_{0_1} X^{{\varepsilon_{\ell'+1,0}}}_{0_0}) \cdot \dots \cdot \\
            & \cdot \text{e}(W'_{\ell-1}, X^{{\varepsilon_{\ell-1 ,2}}}_{0_2} X^{{\varepsilon_{\ell-2,1}}}_{0_1} X^{{\varepsilon_{\ell-1 ,1}}}_{0_1} X^{{\varepsilon_{\ell-1,0}}}_{0_0}) \cdot \\ 
            & \cdot \text{e}(W'_{\ell}, X^{{\varepsilon_{\ell ,2}}}_{0_2} X^{{\varepsilon_{\ell-1,1}}}_{0_1} (C_3)^{{\varepsilon_{\ell ,2}}} (C_4)^{\varepsilon_{\ell-1 ,1}} f^z) \\
            &\text{e}(h^\zeta b^\rho c^\omega {v'}^\tau,{X_0}_0)=\text{e}({v'}^\gamma,X)\\
            &\wedge \ \ \ \ C_1=X^{x_{i^*}}_{0_1}\cdot h^{\alpha} \wedge \ \ \ \ C_2=X^{x_{i^*}}_{0_0}\cdot h^{\beta} \ \ \ \ \wedge \\ 
            &C_3=X^{x_{j^*}}_{0_1}\cdot h^{\delta} \wedge \ \ \ \ C_4=X^{x_{j^*}}_{0_0}\cdot h^{\mu} \bigg\}.
    \end{align*}
    For the provider $A$, $\VVV$ and $\EEE$ denote the vertex and edge sets of multigraph $\GGG'_A$, while $\EEE'$ represents the set of edges along the path, including loops. The same applies to provider $B$ with the corresponding multigraph values.

    The fourth line demonstrates knowledge of the commitments for all vertices of the multigraph, as well as the commitments of all edges not belonging to the path under consideration. Lines five through ten provide the proof of the path's existence.
    Each pairing confirms the presence of the corresponding edge.
    Specifically, the first pairing in line five demonstrates the existence of the first edge, connecting the initial node to the second node of the first edge; the initial node is hidden, which is why the commitments $C_1$ and $C_2$ appear.
    The second pairing confirms the existence of the second edge and that its first node coincides with the second node of the first edge; as ensured by values that will be explained later.
    This reasoning continues up to line ten, where the final edge is reached; its second node, the terminal node, is also hidden by the commitments.
    Starting from line seven, the loop technique comes into play: up to that point, the proof demonstrates the actual path length $\ell'$; in the following lines, it shows the existence of the $\ell-\ell'$ loops, all distinct.
    Line eleven contains the pairing associated with the signature on the multigraph.
    Combined with the products from the preceding lines, it allows verification of equality with respect to the final pairing.
    The last two lines of the proof, on the other hand, demonstrate knowledge of the values that generate the public commitments for the initial and terminal vertices.  
    
    To demonstrate the existence of a path between the public node $i^*$ and a hidden node $j^*$, it would suffice to compute the commitments exclusively for node $j^*$. In the proof, at line 5, in the first pairing $C_1$ and $C_2$ do not appear, instead, $(X_{0_1}^{x_{i^*}})^{\varepsilon_{1,2}}$ and $(X_{0_0}^{x_{i^*}})^{\varepsilon_{1,1}}$ are used. Furthermore, the line referring to the two commitments of the first node (line 12)  is no longer present. Conversely, the same holds when $i^*$ is the hidden node and $j^*$ is the public node.

    Whether providers actually use the announced path relies on the binding (via the commitments described in \ref{subsec:Pedersen} and \ref{subsec: commitment}) and  on the impersonation resilience of the multigraph signature scheme discussed in \ref{subsec:sig}. Any deviation would require either opening the commitments to a different values or impersonating the auditor by producing a valid graph signature on an inconsistent multigraph, both infeasible under the stated computational assumption.

\subsection{Values}

    In this section, we illustrate how the various elements are used, explaining their purpose, theoretical significance, and practical implications.
    
    $$W'_i=a^{o_i r_i^{-1} r \prod_{w \in f_{\VVV(i)}} (x'+w)}_{i_0}, \ \forall i \in \VVV,$$
    
    $$W'_{(i,j)}=a^{o_{(i,j)} r_{(i,j)}^{-1} r (x'+j) \prod_{w \in f_{\EEE{(i,j)}}} (x'+w)}_{{(i,j)}_0}, \ \forall {(i,j)} \in \EEE \setminus \EEE',$$
    
    $$W_k'= a^{o_{(i_k,j_k)} r_{(x_{i^*},x_{j^*})}^{-1} r \prod_{w \in f_{\EEE{(i_k,j_k)}}} (x'+w)}_{{(i_k,j_k)}_0}, \ \text{for} \ k \in \{ 1, \dots ,\ell \},$$
    are public values computed by the prover and are used by the verifier to check the commitment of the graph; note that the verifier is not able to compute those values, since they are derived from graph data that remains hidden. The first one refers to vertices, the second one to edges not belonging to the path and the last one to edges that are part of the path; $\{ r_i, r_{(i,j)}, r_{(x_{i^*},x_{j^*})} \} \in \ZZ^*_p$ are chosen by the prover and constitute secret elements. They are used to randomize each value associated with the various commitments. While $\{ o_i, o_{(i,j)}, o_{(i_k,j_k)}\}$ are the openings of the commitments for the multigraph and $r$ corresponds to the signature's randomization and is used as a component of the exponent of the values $h,b,c$ and $v'.$ Then $\varepsilon_{i_1}=r_i, \ \varepsilon_{i_0}=r_i\cdot i, \ \varepsilon_{(i,j)_1}=r_{(i,j)}, \ \varepsilon_{(i,j)_0}=r_{(i,j)} \cdot i$, and holds that $X_{0_1}^{\varepsilon_{i_1}}X_{0_0}^{\varepsilon_{i_0}}=X_{0_0}^{r_i(x'+i)}$ and $X_{0_1}^{\varepsilon_{{(i,j)}_1}}X_{0_0}^{\varepsilon_{{(i,j)}_0}}=X_{0_0}^{r_{(i,j)}(x'+i)}$.
    The values of $X_{0_0},X_{0_1}$ and $X_{0_2}$ play distinct roles within the proof of the existence of the path (from line 5 to 10).
    $X_{0_0}$ encodes the product of the identifiers of two vertices, $X_{0_1}$ contains the information of a single vertex and, in our case is used to verify that the second node of one edge matches the first node of the subsequent edge.
    Finally, $X_{0_2}$ incorporates the random value $r_{(x_{i^*},x_{j^*})}$ and ties all edges of the path together, assigning the same exponent in $X_{0_2}$ to each edge. 
    In the table \ref{tab:eps-per-edge} are showed the values of $\varepsilon_{i,k}$ along the padded path. Note that $\varepsilon_{i,2}$ represents the common randomization associated to the path, while $\varepsilon_{i,1}$ aligns the nodes of the edges such that the second node of the edge $i$ is the same as the first node of the edge $i+1$.

\begin{table}[H]
\centering
\scriptsize
\setlength{\tabcolsep}{2pt}        
\renewcommand{\arraystretch}{1.1}  
\begin{tabularx}{\columnwidth}{@{}l c *{3}{>{\centering\arraybackslash}X}@{}}
\toprule
\textbf{Edge} & \textbf{\#} & $\boldsymbol{\varepsilon_{i,2}}$ & $\boldsymbol{\varepsilon_{i,1}}$ & $\boldsymbol{\varepsilon_{i,0}}$ \\
\midrule
$(i^*,j_1)$                  & 1      & $r_{(x_{i^*},x_{j^*})}$ & $r_{(x_{i^*},x_{j^*})}j_1$           & -- \\
$(j_1,j_2)$                  & 2      & $r_{(x_{i^*},x_{j^*})}$ & $r_{(x_{i^*},x_{j^*})}j_2$           & $r_{(x_{i^*},x_{j^*})}\, j_1 j_2$ \\
$\cdots$                     & $\cdots$ & $\cdots$               & $\cdots$                             & $\cdots$ \\
$(j_{\ell'-1},j^*)$          & $\ell'$ & $r_{(x_{i^*},x_{j^*})}$ & $r_{(x_{i^*},x_{j^*})} j_{\ell'-1}$  & $r_{(x_{i^*},x_{j^*})} j_{\ell'-1} x_{j^*}$ \\
$(j^*,j^*)$                  & $\ell'+1$ & $r_{(x_{i^*},x_{j^*})}$ & $r_{(x_{i^*},x_{j^*})} x_{j^*}$     & $r_{(x_{i^*},x_{j^*})} x_{j^*}^2$ \\
$\cdots$                     & $\cdots$ & $\cdots$               & $\cdots$                             & $\cdots$ \\
$(j^*,j^*)$                  & $\ell-1$ & $r_{(x_{i^*},x_{j^*})}$ & $r_{(x_{i^*},x_{j^*})} x_{j^*}$     & $r_{(x_{i^*},x_{j^*})} x_{j^*}^2$ \\
$(j^*,j^*)$                  & $\ell$   & $r_{(x_{i^*},x_{j^*})}$ & -     & - \\
\bottomrule
\end{tabularx}
\caption{}
\label{tab:eps-per-edge}
\end{table}
    As can be observed, the values of $\varepsilon_{i,1}$ and $\varepsilon_{i,2}$ are the same for the index $i$ ranging from $\ell'+1$ to $\ell$.
    This occurs because, starting from $\ell'+1$, only loops on the final vertex $j^*$ are used, therefore, the initial and terminal nodes of these edges coincides and both correspond to $j^*$.
    In line 11 $\zeta = sk_U r, \ \rho = s'r, \ \omega = r, \ \tau=ty, \ \gamma=y$ and $v'=v^{ry^{-1}}$.  

\section{Extensions and Variants}
  In the following, we sketch how the basic functionality of proving availability of an inter-network path can be extended to support more expressive statements.
  
\paragraph{Support for Multi-Path Routing}
  An important technique to reduce the trust in the repeater nodes in a QKD network is to use multi-path routing, i.e., sending messages over multiple disjoint paths in a way that the adversary needs to compromise a node on every path to break security, cf.\cite{valbusa_relaxing_2025} for a discussion. This cannot be achieved by the construction presented above, yet can be solved by allowing to additionally encode additional attributes for every node, akin to \cite{cryptoeprint:2023/1181}.
  By adding a random attribute $k_N\in\ZZ_p$ for every node $N$, one can now use techniques similar to Camenisch et al.~\cite{DBLP:conf/sacrypt/CamenischKLMNP15} to derive pseudonyms $nym_N$ of the form $nym_N=H(sid)^{k_N}$ for the given session id $sid$ for every transversed node, and prove consistency of these pseudonyms with the graph signature as part of the zero-knowledge proof. The customer requesting a communication link over $t$ disjoint channels can now simply check that all $t$ connectivity proofs and additionally check whether all pseudonyms are pairwise distinct, thereby obtaining provable guarantees of disjointness.
  
  On the other hand, by the unlinkability of pseudonyms across session ids~\cite{DBLP:conf/sacrypt/CamenischKLMNP15}, the service providers' confidentiality needs remain guaranteed, as no incremental learning of the network topology is possible over multiple sessions.

\paragraph{Support for quality criteria}
  Similar to the work of Krenn et al.~\cite{todo-our-paper}, one could use the same approach as above (i.e., adding attributes to nodes/edges) to assure further quality criteria of the available path.
  That is, by adding attributes (e.g., certification level, bandwidth, manufacturer, ...) to the graph, one could prove availability of one or multiple paths satisfying complex policies, e.g., requiring certification levels or bandwidth over a certain threshold.
  While the actual instantiation is beyond the scope of this paper, this can be achieved using standard techniques~\cite{DBLP:conf/sacrypt/CamenischKLMNP15,DBLP:books/daglib/0039026}.

\section{Conclusions}

We presented a topology-hiding connectivity assurance protocol that enables quantum key distribution (QKD) network providers to jointly prove the existence of secure end-to-end connections without disclosing internal topology information. By extending graph-signature techniques to support multigraphs and committed endpoints, the proposed approach achieves zero-knowledge proofs of connectivity that ensure both soundness and topology hiding. This contribution bridges cryptographic assurance mechanisms with the operational realities of trusted-repeater QKD networks, providing a concrete means to verify inter-network connectivity in a privacy-preserving manner.

Our protocol demonstrates that it is possible to balance verifiability and confidentiality across independently operated QKD domains, offering a path toward auditable, yet topology-protecting, network assurance. The scheme achieves statistical privacy through zero-knowledge techniques while maintaining efficient proof generation and verification, paving the way for scalable certification of both single and multiple disjoint paths in multi-provider QKD scenarios.

\subsection{Future Work}
Several challenges remain open. 
The current construction relies on number-theoretic assumptions for soundness, which limits its post-quantum robustness. 
Designing a variant that maintains statistical privacy while achieving post-quantum soundness represents a critical direction for future work. 

Further research is also required to evaluate performance on realistic network topologies (yet practical efficiency can be expected based on the benchmarks provided in \cite{tan_q-sdh-based_2020}) and formalize auditor governance and key management procedures including, e.g., key rotation.

\ifnum\ANONYMOUS=0
  \section*{Acknowledgments}
  This work has received funding from the European Union's Horizon Europe research and innovation program under No. 101114043 (“QSNP”), from the DIGITAL-2021-QCI-01 Digital European Program under No. 101091642 and the National Foundation for Research, Technology and Development (``QCI-CAT'').
  Views and opinions expressed are however those of the authors only and do not necessarily reflect those of the funding agencies.
  The granting authorities cannot be held responsible for them.
\fi

\bibliographystyle{IEEEtran}
\bibliography{references}

\end{document}